\documentclass[12pt]{article}

\usepackage{amsmath,amssymb,amsthm,amsfonts,mathtools}
\usepackage[margin=1in]{geometry}
\usepackage[hidelinks]{hyperref}
\usepackage[nameinlink,capitalize]{cleveref}
\usepackage{setspace}
\usepackage{booktabs}
\usepackage{floatrow}
\usepackage{enumerate}
\usepackage{natbib}
\usepackage{xurl}
\newtheorem{theorem}{Theorem}
\newtheorem{definition}{Definition}
\newtheorem{assumption}{Assumption}
\newtheorem{lemma}{Lemma}
\newtheorem{proposition}{Proposition}
\newtheorem{remark}{Remark}

\begin{document}
	
	\title{Performance Rating Equilibrium}
	\author{Mehmet Mars Seven\thanks{Department of Political Economy, King's College London, UK. E-mail: mehmetmarsseven@gmail.com}}
	\date{\today}
	
	\maketitle
	
	\begin{abstract}
		In this note, we introduce a novel performance rating system called Performance Rating Equilibrium (PRE). A PRE is a vector of ratings for each player, such that if these ratings were each player's initial rating at the start of a tournament, scoring the same points against the same opponents would leave each player's initial rating unchanged. In other words, all players' initial ratings perfectly predict their actual scores in the tournament. This property, however, does not hold for the well-known Tournament Performance Rating. PRE is defined as a fixed point of a multidimensional rating function.  We show that such a fixed point, and hence a PRE, exists under mild conditions. We provide an implementation of PRE along with several empirical applications. PREs have broad applicability, from sports competitions to the evaluation of large language models.
	\end{abstract}
	
	
	\section{Introduction}
	
	Rating and ranking systems are widely studied across various fields, including (political) economics, social choice, sports, and computer science, to assess individual performance or rank objects. One prominent approach, the Elo rating system, originally developed for chess by \citet{elo1978}, has been applied in diverse domains, including association football \citep{FIFA, csato2021}, tennis \citep{Williams2021}, college rankings \citep{Avery2013}, dating applications \citep{Kosoff2016}, and large language model evaluation \citep{zheng2023}.\footnote{Alternative systems, such as the Glicko rating system \citep{Glickman1995}, may also be used for ratings in these areas.} 
	
	In the Elo rating system, players are assigned a numerical rating, which is updated based on their performance over a given period. A player gains Elo rating if his or her score against opponents is better than the expected score predicted by the Elo ratings. Since Elo rating is cumulative, based on all games that a player has played, it does not necessarily reflect a player's recent performance. The Tournament Performance Rating (TPR) is the standard method for calculating a player's performance in a subset of games, such as a specific tournament.\footnote{In the case of LLM evaluation, such as in the Chatbot Arena, the TPR of an LLM can be calculated based on a single-day performance or a subject-specific performance, such as mathematics.} A player's TPR is the hypothetical rating $R$ that would remain unchanged based on the player's performance in a tournament. However, because TPR is calculated directly from opponents' pre-tournament ratings, it does not take into account their actual performances in the event.
	
	Here, we introduce the concept of \textit{Performance Rating Equilibrium} (PRE), represented by a vector $x^*$ of hypothetical ratings. A PRE is defined as a fixed point of a multidimensional rating function: if each player $i$ began the tournament with an initial rating $x^*_i$, then achieving the same points against the same opponents would leave each player's initial rating $x^*_i$ unchanged. In other words, these initial ratings perfectly predict each player's final score, and we refer to each player's rating at a PRE as a perfect performance rating (PPR).\footnote{The term `equilibrium' here refers to an equilibrium point or steady state of a dynamical system. This concept is not choice-based, though it is interactive, as each PPR in a PRE depends on the other PPRs.}  
	
	\citet{landau1895,landau1914} introduced one of the earliest score-based ranking systems  inspired by round-robin chess tournaments. Iterative methods for ranking, whether of players or political candidates, date back to at least \citet{wei1952,kendall1955}, who extended an approach by \citet{wei1952}. For some selected further developments, see \citet{keener1993,laslier1997,herings2005,brozos2008,gonzalez2014,csato2015, csato2017}, as well as the influential PageRank algorithm  introduced by \citet{brin1998}. \citet{herings2005} propose a new method for ranking the nodes in directed graphs, while \citet{gonzalez2014} provide an axiomatic analysis of several ranking methods, including generalised row sum methods by \citet{chebotarev1994}.\footnote{\citet{rubinstein1980b} provides an axiomatization of (non-iterative) score-based ranking.} In addition, \citet{csato2017} explores Swiss-system team tournaments, proposing various score-based ranking methods essentially based on generalised row sum methods and applying them to the European Chess Team Championships.\footnote{For a broader, growing literature on fairness in sports, see, e.g., \citet{brams2018,brams2018b,csato2021,Lambers2021,Palacios2023,krumer2023,csato2024,devriesere2024}, and the references therein.}

	The early pioneering work by \citet{kendall1955} focused on recursive rankings that incorporate both the players' current performances (measured by scores) and the performances of their opponents in the tournament. To my knowledge, the study by \citet{brozos2008} is the first to use performance ratings, called ``Recursive Performance,'' that account for the opponents' current performance ratings rather than only current scores in their ranking method. There are three main differences between Recursive Performance and the current study. First, Brozos-Vázquez et al.'s main focus is on ranking participants, whereas the focus here is on finding ratings that perfectly predict the actual performances of players. Second, although Recursive Performance offers an improvement over the standard TPR, it, like TPR, does not generally yield a fixed point of the rating function, so these concepts are not in equilibrium as defined here. Lastly, Recursive Performance ratings are calculated using average opponent ratings with a corrective factor based on score. Although this approach approximates actual performance ratings reasonably well when rating gaps are small, it is well known that it can lead to paradoxical outcomes, such as a player's performance rating decreasing after winning against a low-rated opponent.
	
	\section{Setup}
	
	\noindent
	Let $T=(N,Y,W,f,\psi)$ denote a \emph{tournament}, where:
	\begin{enumerate}
		\item $N=\{1,2,\dots,n\}$ is the set of players.
		
		\item $W=(w_{ij})_{i,j\in N}$ is the symmetric \emph{schedule matrix}, where $w_{ij}=w_{ji}\ge 0$ denotes the number (or weight) of games between players $i$ and $j$, and $w_{ii}=0$.
		
		\item $Y=(y_{ij})_{i,j\in N}$ with $0\le y_{ij}\le w_{ij}$ denotes realized points earned by $i$ against $j$, with
		\[
		y_{ij}+y_{ji}=w_{ij}\quad\text{for all }i\ne j.
		\]
		
		\item $k\in\mathbb{R}_{\ge 0}^n$ is the vector of \emph{games played} by each player, derived from $W$:
		\[
		k_i:=\sum_{j\neq i} w_{ij}.
		\]
		
		\item $m\in\mathbb{R}_{\ge 0}^n$ is the vector of \emph{scores}, derived from $Y$:
		\[
		m_i:=\sum_{j\ne i} y_{ij}.
		\]
		The total score is
		\[
		M := \sum_{i=1}^n m_i \;=\; \sum_{i<j} w_{ij}.
		\]
		We assume $0<m_i<k_i$ for all $i$ with $k_i>0$.
		
		\item $f:\mathbb{R}^n\to\mathbb{R}^n$ is the \emph{performance rating function} associated with $T$. For any hypothetical initial ratings $x\in\mathbb{R}^n$, the vector $f(x)$ returns the performance ratings implied by $T$.
		
		\item Finally, $\psi:\mathbb{R}\to(0,1)$ (defined below) is a monotone antisymmetric function that gives the win probability (or the expected score share) in a single game as a function of the rating difference.
	\end{enumerate}
	
	\begin{definition}[Win Probability]
		\label{def:win_probability}
		The win probability function $\psi:\mathbb{R}\to(0,1)$ models the expected score share of a player in a single game against an opponent, depending on the difference in their ratings. We assume:
		\begin{enumerate}[(a)]
			\item $\psi$ is $C^1$ and strictly increasing: $\psi'(t)>0$ for all $t$;
			\item $\psi(-t)=1-\psi(t)$ for all $t$ (antisymmetry);
			\item $\lim_{t\to+\infty}\psi(t)=1$ and $\lim_{t\to-\infty}\psi(t)=0$.
		\end{enumerate}
	\end{definition}

	\begin{definition}[Performance Rating Equilibrium]\label{def:pre-fixedpoint}
		Given a tournament $T$, a vector $x^*\in\mathbb{R}^n$ is a \emph{Performance Rating Equilibrium (PRE)} if it is a fixed point of the performance rating function $f$:
		\[
		f(x^*)=x^*.
		\]
		We call $x_i^*$ the \emph{perfect performance rating (PPR)} of player $i$ at the PRE.
	\end{definition}

	\begin{definition}[Expected Score]\label{def:exp_score}
		Given $x\in\mathbb{R}^n$, write $\Delta_{ij}(x):=(x_i-x_j)/\beta$, where $\beta>0$, and define the model-implied expected (total) score of player $i\in N$ in tournament $T$ by
		\[
		E_i(x):=\sum_{j\ne i} w_{ij}\,\psi\big(\Delta_{ij}(x)\big).
		\]
		Let $E(x)=(E_i(x))_{i\in N}$.
	\end{definition}
	
	\begin{remark}\label{rem:even-derivative}
		Antisymmetry and differentiability imply $\psi'(-t)=\psi'(t)$, so $\psi'$ is even. Moreover, only rating differences matter: for any $c\in\mathbb{R}$, $E(x+c\mathbf{1})=E(x)$, where $\mathbf{1}$ the vector of all ones.
	\end{remark}

	\begin{lemma}[PRE equivalence]\label{lem:pre-equivalence}
		There exists a choice of $f$, the canonical $\psi$–performance function, such that
		\[
		f(x)=x \quad\Longleftrightarrow\quad E(x)=m.
		\]
		Therefore, the PRE condition is exactly $E(x^*)=m$.
	\end{lemma}
	
	\begin{proof}
		Fix a tournament $T$ and realized score vector $m=(m_i)_{i\in N}$.
		For each player $i\in N$, define the component function $f_i:\mathbb{R}^n\to\mathbb{R}$ by the implicit equation
		\begin{equation}\label{eq:fi-def}
			\sum_{j\neq i} w_{ij}\,\psi\Big(\frac{f_i(x)-x_j}{\beta}\Big) = m_i.
		\end{equation}
		Because $\psi$ is continuous, strictly increasing, and satisfies 
		$\lim_{t\to-\infty}\psi(t)=0$ and $\lim_{t\to\infty}\psi(t)=1$ by Definition~\ref{def:win_probability},
		the left-hand side of \eqref{eq:fi-def} is continuous and strictly increasing in the unknown $z=f_i(x)$,
		with limits $0$ and $\sum_{j\neq i}w_{ij}=k_i$ as $z\to-\infty$ and $z\to\infty$, respectively.
		Since $m_i\in (0,k_i)$, the intermediate value theorem ensures the existence of a unique solution $z=f_i(x)$.
		
		Collect these component functions into the vector function
		\[
		f(x) := \big(f_1(x),\dots,f_n(x)\big)^\top:\mathbb{R}^n\to\mathbb{R}^n,
		\]
		which we call the \emph{canonical $\psi$–performance function}.
		
		Now we show the equivalence.
		
		($\Rightarrow$) Suppose $f(x)=x$. Then for each $i\in N$, substituting $f_i(x)=x_i$ into \eqref{eq:fi-def} yields
		\[
		\sum_{j\neq i} w_{ij}\,\psi\Big(\frac{x_i - x_j}{\beta}\Big) = m_i,
		\]
		that is, $E_i(x)=m_i$. Hence $E(x)=m$.
		
		($\Leftarrow$) Conversely, suppose $E(x)=m$. Then for each $i\in N$,
		\[
		\sum_{j\neq i} w_{ij}\,\psi\Big(\frac{x_i - x_j}{\beta}\Big) = m_i,
		\]
		so $x_i$ satisfies \eqref{eq:fi-def}. Because the solution to \eqref{eq:fi-def} is unique, we have $f_i(x)=x_i$ for all $i$. Therefore $f(x)=x$.
		
		Thus $f(x)=x$ if and only if $E(x)=m$.  In particular, a fixed point $x^*$ of $f$ satisfies $E(x^*)=m$, which is exactly the defining condition of a PRE.
	\end{proof}
	
	\noindent
	In Section~\ref{sec:characterization}, we characterize the PRE as the maximizer of a concave potential.

	\subsection{The standard Elo model}
	
	Here, we note that the standard Elo system corresponds to a case in which $\psi$ is a logistic function.
	
	\begin{proposition}\label{prop:elo_is_psi}
		The standard Elo win probability function, defined with a scaling factor of 400 and $\beta=1$ as
		\begin{equation}\label{eq:elo_psi}
			\psi_{\text{Elo}}(t) = \frac{1}{1 + 10^{-t/400}},
		\end{equation}
		satisfies the properties of a win probability function as per Definition~\ref{def:win_probability}.
	\end{proposition}
	\begin{proof}
		We verify the required properties for $\psi(t)=\psi_{\text{Elo}}(t)$. Let $C=400$.
		In terms of domain and range, $\psi(t)$ is defined for all $t\in\mathbb{R}$. Since $10^{-t/C}>0$, we have $1+10^{-t/C}>1$, so $0 < \psi(t) < 1$.
		
		As for smoothness, $\psi(t)$ is composed of elementary functions and is therefore $C^1$. 
		
		For monotonicity, we calculate the derivative.
		\begin{align*}
			\psi'(t) &= \frac{d}{dt} (1+10^{-t/C})^{-1} \\
			&= -(1+10^{-t/C})^{-2} \cdot (10^{-t/C} \cdot \ln(10) \cdot (-1/C)) \\
			&= \frac{\ln(10)}{C} \frac{10^{-t/C}}{(1+10^{-t/C})^2}.
		\end{align*}
		Since $\ln(10)>0, C>0$, and the exponential terms are positive, $\psi'(t)>0$. Thus, $\psi(t)$ is strictly increasing.
		
		As for antisymmetry, we check if $\psi(-t)=1-\psi(t)$.
		\begin{align*}
			1-\psi(t) &= 1 - \frac{1}{1+10^{-t/C}} = \frac{10^{-t/C}}{1+10^{-t/C}}.
		\end{align*}
		Multiplying the numerator and denominator by $10^{t/C}$:
		\[
		1-\psi(t) = \frac{1}{10^{t/C}+1} = \psi(-t).
		\]
		Finally, as $t\to+\infty$, $-t/C\to-\infty$, so $10^{-t/C}\to 0$, and $\psi(t)\to 1$. As $t\to-\infty$, $-t/C\to+\infty$, so $10^{-t/C}\to \infty$, and $\psi(t)\to 0$.
	\end{proof}
	
	A related concept in Elo systems is the Tournament Performance Rating (TPR), which is a specific type of performance rating function. Given opponents' pre-tournament ratings, the TPR$_i$ of player $i$ is defined as the rating level that would have predicted the observed score $m_i$ with respect to $i$'s opponents' \textit{pre-tournament ratings}, as opposed to the PRE.
	
	\begin{definition}[Tournament Performance Rating]
		Assume that player $i$ plays $k_i$ games against opponents with ratings $x^{(1)}, \dots, x^{(k_i)}$. Then, TPR$_i$ satisfies:
		\begin{equation}
			\label{eq:tpr_definition}
			m_i = \sum_{j=1}^{k_i} \psi_{\text{Elo}}(\text{TPR}_i - x^{(j)}).
		\end{equation}
	\end{definition}
	
	As Elo noted \citep[p. 12]{elo1978}, TPR is undefined for zero or perfect scores.\footnote{\citet{ismail2023} introduces Complete Performance Rating (CPR), which is defined for all scores, including when $m_i=0$ or $m_i=k_i$.}
	
	Note also that the PRE condition $E(x^*)=m$ corresponds to finding $x^*$ such that for all $i$:
	\[
	m_i = \sum_{j\ne i} \frac{1}{1 + 10^{(x^*_j - x^*_i)/400}}.
	\]
	Thus, unlike TPR, PRE does not depend on the pre-tournament ratings and it corresponds to a fixed point of the performance function.
	
	\section{Characterization and uniqueness}\label{sec:characterization}
	
	We characterize the PRE as the solution to a concave maximization problem. This requires assumptions on the tournament structure and outcomes.
	
	\begin{assumption}[Irreducibility]\label{A1}
		$W$ is irreducible; i.e., $N$ cannot be partitioned into non-empty disjoint sets $S_1, S_2$ such that $w_{ij}=0$ for all $i\in S_1, j\in S_2$.
	\end{assumption}
	
	Irreducibility means that the tournament graph is strongly connected.
	
	\begin{assumption}[Non-dominance]\label{A2}
		For every non-empty proper subset $S\subset N$, if the total interaction weight across the partition $W(S):=\sum_{i\in S,\,j\notin S}w_{ij}>0$, then the outcomes across the partition are non-extreme:
		\[
		\bar{Y}(S^c):=\sum_{i\in S,\,j\notin S}y_{ji}>0\quad\text{and}\quad
		\bar{Y}(S):=\sum_{i\in S,\,j\notin S}y_{ij}>0.
		\]
	\end{assumption}
	Non-dominance means that across any split of players who played each other, neither side achieved a 100\% score against the other.
	
	We now construct a potential function $P(x)$. Let $A(t)$ be the antiderivative of $\psi(t)$ normalized such that $A(0)=0$:
	\[
	A(t) \;=\; \int_0^t \psi(s)\,ds.
	\]
	The antisymmetry of $\psi$ implies $A(t)$ satisfies $A(t)-A(-t)=t$.
	
	We define a symmetric pairwise potential function $G:\mathbb{R}^2\to\mathbb{R}$:
	\[
	G(a,b) := \beta A\left(\frac{a-b}{\beta}\right) + b.
	\]
	$G$ is convex since $A(t)$ is convex, as its derivative $\psi(t)$ is increasing. Indeed, it is symmetric: $G(a,b)=G(b,a)$.
	$G(b,a) = \beta A(\frac{b-a}{\beta}) + a$.
	$G(a,b)-G(b,a) = \beta(A(\frac{a-b}{\beta})-A(\frac{b-a}{\beta})) + b-a$.
	Let $t=(a-b)/\beta$. The expression becomes $\beta(A(t)-A(-t)) - \beta t = \beta t - \beta t = 0$.
	
	Its derivatives are
	$\partial_a G(a,b) = \beta \psi(\frac{a-b}{\beta})\frac{1}{\beta} = \psi(\Delta_{ab})$, where $\Delta_{ab}:=(a-b)/\beta$.
	$\partial_b G(a,b) = \beta \psi(\frac{a-b}{\beta})\frac{-1}{\beta} + 1 = 1-\psi(\Delta_{ab}) = \psi(\Delta_{ba})$.
	
	Finally, it satisfies $G(a+c, b+c)=G(a,b)+c$.

	Next, define the convex regularizer $R(x)$ and the concave potential function $P:\mathbb{R}^n\to\mathbb{R}$ as:
	\[
	R(x) := \sum_{i<j}w_{ij}\,G(x_i, x_j),
	\]
	and 
	\[
	P(x) := m^\top x - R(x),
	\]
	respectively.

	\begin{proposition}\label{prop:concavity}
		$P(x)$ is concave and translation invariant, $P(x+c\mathbf{1})=P(x)$. A vector $x^*$ is a PRE if and only if $\nabla P(x^*)=0$.
	\end{proposition}

	\begin{proof}
		We first compute the gradient of $R(x)$. For $k\in N$,
		\begin{align*}
			\frac{\partial R}{\partial x_k}
			&= \sum_{j>k} w_{kj}\, \partial_1 G(x_k, x_j) \;+\; \sum_{i<k} w_{ik}\, \partial_2 G(x_i, x_k) \\
			&= \sum_{j>k} w_{kj}\, \psi\!\big(\Delta_{kj}(x)\big) \;+\; \sum_{i<k} w_{ki}\, \psi\!\big(\Delta_{ki}(x)\big) \\
			&= \sum_{j\ne k} w_{kj}\, \psi\!\big(\Delta_{kj}(x)\big) \;=\; E_k(x).
		\end{align*}
		Thus $\nabla P(x)=m-E(x)$. In particular, the first–order condition $\nabla P(x^*)=0$ is exactly the PRE condition $E(x^*)=m$. Since $P$ is concave (see below), any $x^*$ with $\nabla P(x^*)=0$ is a global maximizer of $P$, and any maximizer satisfies $\nabla P(x^*)=0$.
		
		Summing the expectations over all players and using $\psi\!\big(\Delta_{ij}(x)\big)+\psi\!\big(\Delta_{ji}(x)\big)=1$ yields
		\[
		\sum_{i=1}^n E_i(x)
		= \sum_{i<j} w_{ij}\Big(\psi\!\big(\Delta_{ij}(x)\big)+\psi\!\big(\Delta_{ji}(x)\big)\Big)
		= \sum_{i<j} w_{ij} = M
		= \sum_{i=1}^n m_i.
		\]
		Hence $\mathbf{1}^\top\nabla P(x)=\sum_i(m_i-E_i(x))=0$ for all $x$. In particular, at any maximizer of $P$ over $H=\{x:\sum_i x_i=0\}$, the Lagrange multiplier must be $0$.
		
		The Hessian $\nabla^2 P(x)$ is the negative Jacobian of $E(x)$, and its quadratic form is
		\begin{equation}\label{eq:Hessian}
			h^\top \nabla^2 P(x)\, h \;=\; -\frac{1}{\beta}\sum_{i<j} w_{ij}\,\psi'\!\big(\Delta_{ij}(x)\big)\,(h_i-h_j)^2.
		\end{equation}
		Since $w_{ij}\ge 0$ and $\psi'(t)>0$ for all $t$, the Hessian is negative semidefinite, so $P$ is concave.
		
		For translation invariance, using $G(a+c,b+c)=G(a,b)+c$ we obtain
		\[
		R(x+c\mathbf{1})=\sum_{i<j} w_{ij}\big(G(x_i,x_j)+c\big)=R(x)+c\,M,
		\]
		and therefore
		\[
		P(x+c\mathbf{1})=m^\top(x+c\mathbf{1})-R(x+c\mathbf{1})=m^\top x+c\sum_i m_i-(R(x)+cM)=P(x),
		\]
		since $\sum_i m_i=M$. 
	\end{proof}

	\subsection{Existence}
	Existence requires that the maximum of $P(x)$ is attained. Since $P(x)$ is translation invariant, we analyze its coercivity on the subspace $H=\{x\in\mathbb{R}^n:\sum x_i=0\}$. $P|_H$ is coercive if $P(x)\to-\infty$ as $\|x\|\to\infty$, $x\in H$.

	\begin{theorem}[Existence]\label{thm:existence}
		A finite PRE exists if and only if Non-dominance (Assumption~\ref{A2}) holds within each irreducible component of $W$.
	\end{theorem}
	
	\begin{proof}
		Because $P$ is translation-invariant, we restrict to $H=\{x:\sum_i x_i=0\}$ and study the one-dimensional profiles $t\mapsto P(x_0+th)$ along rays with $h\in H\setminus\{0\}$. We (i) rewrite the directional derivative $D_hP$ in a pairwise form, (ii) define and compute its recession limit $L(h)$, (iii) deduce that $L(h)<0$ for all $h\ne 0$ implies coercivity and hence existence, and (iv) show that failure of Non-dominance produces a direction with $D_hP>0$ for all $t\ge 0$, so no maximizer exists.
		
		Step 1: Pairwise form of the directional derivative.
		Assume $W$ is irreducible (\cref{A1}); otherwise the argument applies component-wise.\footnote{If $W$ is reducible, $P$ decomposes as a sum over components and the analysis applies on each component independently, yielding existence if and only if \cref{A2} holds on every component.}
		For $h\in H\setminus\{0\}$, the directional derivative is $D_h P(x)=h^\top\nabla P(x)=h^\top(m-E(x))$. Grouping terms by unordered pairs $\{i,j\}$ gives
		\begin{align}\label{eq:dir_deriv}
			D_h P(x)
			&= \sum_i h_i m_i - \sum_i h_i E_i(x) \nonumber \\
			&= \sum_{i\ne j} h_i y_{ij} - \sum_{i\ne j} h_i w_{ij}\psi(\Delta_{ij}(x)) \nonumber \\
			&= \sum_{i<j} (h_i y_{ij} + h_j y_{ji}) - \sum_{i<j} w_{ij}\big(h_i\psi(\Delta_{ij}) + h_j\psi(\Delta_{ji})\big) \nonumber \\
			&= \sum_{i<j} (h_i-h_j)\big(y_{ij}-w_{ij}\psi(\Delta_{ij}(x))\big),
		\end{align}
		using $y_{ij}+y_{ji}=w_{ij}$ and $\psi(\Delta_{ij})+\psi(\Delta_{ji})=1$.
		
		Step 2: The recession limit and coercivity.
		For a fixed base point $x_0\in H$, define the ray $x_t:=x_0+th$ and the recession limit
		\[
		L(h):=\lim_{t\to\infty} D_h P(x_t).
		\]
		This limit exists and is independent of $x_0$, since
		$\Delta_{ij}(x_t)=\Delta_{ij}(x_0)+\frac{t}{\beta}(h_i-h_j)$.
		If $L(h)<0$ for every $h\in H\setminus\{0\}$, then $P|_H$ is coercive: indeed, by the definition of $L(h)$ there exist $\varepsilon>0$ and $T<\infty$ with
		$D_hP(x_t)\le -\varepsilon$ for all $t\ge T$, hence by the fundamental theorem of calculus along the $C^1$ curve $t\mapsto x_t$,
		\[
		P(x_t)\le P(x_T)-\varepsilon\,(t-T)\xrightarrow[t\to\infty]{}-\infty.
		\]
		
		Step 3: Compute $L(h)\le 0$.
		From \eqref{eq:dir_deriv} and $\Delta_{ij}(x_t)=\Delta_{ij}(x_0)+\tfrac{t}{\beta}(h_i-h_j)$, as $t\to\infty$:
		if $h_i>h_j$, then $\psi(\Delta_{ij}(x_t))\to 1$ and the pair $\{i,j\}$ contributes $(h_i-h_j)(y_{ij}-w_{ij})=-(h_i-h_j)y_{ji}$;
		if $h_i<h_j$, then $\psi(\Delta_{ij}(x_t))\to 0$ and the contribution is $(h_i-h_j)y_{ij}=-(h_j-h_i)y_{ij}$.
		Equivalently, indexing each unordered pair once by the orientation with $h_i>h_j$, we obtain
		\[
		L(h) \;=\; -\sum_{\substack{i\ne j\\ h_i>h_j}} (h_i-h_j)\,y_{ji}\ \le\ 0.
		\]
		
		Step 4: Non-dominance $\Rightarrow$ existence.
		Assume \cref{A2} holds. If there were $h\in H\setminus\{0\}$ with $L(h)=0$, then each summand in the equation above must vanish, so $y_{ji}=0$ whenever $h_i>h_j$.
		Let $S=\{i:\,h_i=\max_\ell h_\ell\}$; since $h\in H$ and $h\ne 0$, $S$ is a nonempty proper subset.
		For every $i\in S$ and $j\notin S$, we have $h_i>h_j$, hence $y_{ji}=0$ and so $\bar{Y}(S^c)=\sum_{i\in S,\,j\notin S}y_{ji}=0$.
		By irreducibility (\cref{A1}), $W(S)>0$ (otherwise $S$ would be disconnected), contradicting Non-dominance (\cref{A2}).
		Thus $L(h)<0$ for all $h\ne 0$, so by Step~2, $P|_H$ is coercive and (since $P$ is continuous) attains a maximizer $x^*\in H$.
		At any maximizer under the linear constraint $\sum_i x_i=0$, the KKT condition gives $\nabla P(x^*)=\lambda\mathbf{1}$; using $\mathbf{1}^\top\nabla P(x)=0$ for all $x$ (mass conservation), we get $\lambda=0$, hence $\nabla P(x^*)=0$ and therefore $E(x^*)=m$.
		
		Step 5: Failure of Non-dominance $\Rightarrow$ nonexistence.
		If \cref{A2} fails, there exists a nonempty proper $S\subset N$ with $W(S)>0$ and, say, a sweep by $S$: $y_{ij}=w_{ij}$ for all $i\in S,\,j\notin S$ (equivalently $y_{ji}=0$).
		Fix $x_0\in H$ and set $x_t:=x_0+th$, where
		\[
		h_i=\frac{1}{|S|}\ (i\in S),\qquad h_j=-\frac{1}{|S^c|}\ (j\notin S).
		\]
		Then $h\in H$, $h_i-h_j=0$ within $S$ and within $S^c$, and $h_i-h_j=\frac{1}{|S|}+\frac{1}{|S^c|}>0$ across the cut.
		Using \eqref{eq:dir_deriv} and the identities $y_{ji}=w_{ij}-y_{ij}$ and $\psi(\Delta_{ji})=1-\psi(\Delta_{ij})$, one may index each cross pair by $(i\in S,j\notin S)$ and obtain
		\[
		D_h P(x_t)=\sum_{i\in S,\;j\notin S}(h_i-h_j)\Big(y_{ij}-w_{ij}\psi\big(\Delta_{ij}(x_t)\big)\Big).
		\]
		Under the sweep assumption $y_{ij}=w_{ij}$, this becomes
		\[
		D_h P(x_t)=\Big(\tfrac{1}{|S|}+\tfrac{1}{|S^c|}\Big)\sum_{i\in S,\;j\notin S} w_{ij}\,\Big(1-\psi\big(\Delta_{ij}(x_t)\big)\Big).
		\]
		Since $W(S)=\sum_{i\in S,\,j\notin S}w_{ij}>0$ and $\psi:\mathbb{R}\to(0,1)$, each summand is positive for all finite $t$; hence $D_hP(x_t)>0$ for all $t\ge 0$.
		Therefore $t\mapsto P(x_t)$ is strictly increasing on $[0,\infty)$, so $P$ has no maximizer on $H$ and no finite PRE exists.
		
		Combining Steps 4 and 5 yields the stated equivalence.
	\end{proof}
	
	\subsection{Uniqueness}
	Uniqueness depends on the strict concavity of $P(x)$, governed by the null space of the Hessian $\nabla^2 P(x)$.
	
	\begin{theorem}[Uniqueness]\label{thm:uniqueness}
		Under Irreducibility (Assumption~\ref{A1}), if a PRE exists, it is unique in $H$. The set of all PREs in $\mathbb{R}^n$ is the line $\{x^*+c\mathbf{1}: c\in\mathbb{R}\}$.
	\end{theorem}
	\begin{proof}
		From \eqref{eq:Hessian}, since $\psi'(t)>0$ everywhere, $h^\top\nabla^2 P(x)\,h = 0$ if and only if $(h_i-h_j)=0$ whenever $w_{ij}>0$. If $W$ is irreducible (the graph is connected), this implies $h_i=h_j$ for all $i,j$. Thus, the null space of $\nabla^2 P(x)$ is exactly $\mathrm{span}(\mathbf{1})$.
		
		For any $h\in H, h\ne 0$, we have $h\notin \mathrm{span}(\mathbf{1})$. Therefore, $h^\top\nabla^2 P(x) h < 0$. This means $P|_H$ (the restriction of $P$ to $H$) is strictly concave. A strictly concave function can have at most one maximizer. If a PRE exists (Theorem~\ref{thm:existence}), the unique maximizer $x^*\in H$ is the unique PRE in $H$. The translation invariance of $P(x)$ (Proposition~\ref{prop:concavity}) implies the set of all maximizers in $\mathbb{R}^n$ is $\{x^*+c\mathbf{1}\}$.
	\end{proof}
	
	\section{Applications}
	
	There may be multiple PREs due to translation invariance. However, there are various ways in which a PRE can be selected. In the examples below, we numerically calculate the fixed point that converges from an initial vector in which all players have the tournament average rating.\footnote{All ratings in the examples are rounded.} While equilibrium selection can also be done using other methods, such as starting with the initial ratings vector instead of the average ratings vector, a PRE obtained with the latter approach has the advantage of not depending on the specific ratings that players had at the start of the tournament.\footnote{To be sure, a PRE is generally independent of the initial ratings; though this information can be used for equilibrium selection.} In other words, this approach is entirely performance-based, given the tournament average rating. A numerical approximation for calculating PREs is demonstrated in the following implementation:
	\noindent \url{https://github.com/drmehmetismail/Perfect-Performance-Ratings}.\footnote{FIDE (International Chess Federation) uses certain approximations and specific rules when updating ratings and calculating performance ratings. For a more detailed discussion, see \citet{ismail2023}, who gives a definition of performance ratings for all possible scores.}

	Table~\ref{tab:ppr1} presents the PPRs and TPRs for two different three-player round-robin tournaments. Let the 2450-rated player be denoted as A, the 2200-rated player as B, and the 2000-rated player as C. In Tournament 1, the final ranking is the reverse of the initial ranking, indicating that the pre-tournament ratings do not accurately reflect the players' actual performances. As a result, there is a stark contrast between the TPRs and PPRs. Here, player C, who is underrated, scores 1.5 points in two games, while player A, who is overrated, scores only 0.5 points. According to Equation~\ref{eq:elo_psi}, the expected score of a 2348-rated player (C) against players rated 2217 and 2085 is 1.5. Similarly, the expected score of a 2085-rated player (A) against players rated 2217 and 2348 is 0.5. 
	
	In summary, the PPRs are set such that players' expected scores, based on their opponents' PPRs, equal their actual points in the tournament. These values are obtained by numerically calculating the fixed point, converged from the initial vector of ratings (2219, 2219, 2219), where 2219 is the tournament's average rating.\footnote{Convergence to the fixed point was achieved after eight iterations.}
	
	\begin{table}[]
		\centering
		\small
		\begin{tabular}{llll|llll} 
			\multicolumn{4}{c}{Tournament 1} & \multicolumn{4}{c}{Tournament 2} \\ 
			\toprule 
			Rating & Points & TPR & PPR & Rating & Points & TPR & PPR   \\ 
			\midrule 
			2000 & 1.5 & 2538 & 2348 & 2450 & 1.5 & 2305 & 2348   \\ 
			2200 & 1.0 & 2225 & 2217 & 2000 & 1.0 & 2325 & 2217  \\ 
			2450 & 0.5 & 1895 & 2085 & 2200 & 0.5 & 1961 & 2085 \\ 
			\bottomrule 
		\end{tabular}
		\caption{Perfect Performance Ratings in three-player round-robin tournaments with different outcomes}
		\label{tab:ppr1}
		\floatfoot{Note: TPR = Tournament Performance Rating; PPR = Perfect Performance Rating}
	\end{table}
	
	In Tournament 2, there are also significant differences between the TPRs and PPRs. Notably, player C's TPR is higher than player A's TPR, even though C scores fewer points than A. This discrepancy arises in part because player A is overrated; that is, A's pre-tournament rating is too high relative to A's final score. For player B, TPR is lower than PPR, as the PPR reflects the strong performance of player C, who is significantly underrated. As this table illustrates, PPRs are entirely based on actual tournament performance and the tournament average rating, without being affected by specific pre-tournament ratings.
	
	\begin{table}[h!]
		\centering
		\small
		\begin{tabular}{lllll|lllll}
			\multicolumn{5}{c}{Palma de Mallorca} & \multicolumn{5}{c}{Sharjah} \\
			\toprule
			Name           & Rtg  & Pts & TPR  & PPR  & Name            & Rtg  & Pts & TPR  & PPR  \\
			\midrule
			Aronian        & 2801 & 5.5  & 2821 & 2857 & Vachier-Lagrave             & 2796 & 5.5  & 2823   & 2860 \\
			Jakovenko      & 2721 & 5.5  & 2824 & 2840 & Grischuk        & 2742 & 5.5  & 2828   & 2852 \\
			Nakamura       & 2780 & 5    & 2788 & 2830 & Mamedyarov      & 2766 & 5.5  & 2813   & 2851 \\
			Svidler        & 2763 & 5    & 2779 & 2815 & Nepomniachtchi  & 2749 & 5.0  & 2764   & 2795 \\
			Tomashevsky    & 2702 & 5    & 2788 & 2813 & Nakamura        & 2785 & 5.0  & 2776   & 2812 \\
			Harikrishna    & 2738 & 5    & 2764 & 2789 & Jakovenko       & 2709 & 5.0  & 2781   & 2818 \\
			Ding Liren     & 2774 & 5    & 2768 & 2783 & Adams           & 2751 & 5.0  & 2776   & 2795 \\
			Rapport        & 2692 & 5    & 2758 & 2743 & Ding            & 2760 & 5.0  & 2748   & 2753 \\
			Radjabov       & 2741 & 5    & 2760 & 2743 & Eljanov         & 2759 & 4.5  & 2693   & 2675 \\
			Vachier-Lagrave            & 2796 & 4.5  & 2741 & 2768 & Rapport         & 2692 & 4.5  & 2726   & 2694 \\
			Eljanov        & 2707 & 4.5  & 2724 & 2706 & Li Chao         & 2720 & 4.5  & 2722   & 2700 \\
			Inarkiev       & 2683 & 4.5  & 2735 & 2699 & Vallejo Pons    & 2709 & 4.5  & 2714   & 2688 \\
			Giri           & 2762 & 4    & 2696 & 2695 & Hou Yifan       & 2651 & 4.0  & 2685   & 2689 \\
			Vallejo Pons   & 2705 & 4    & 2682 & 2643 & Aronian         & 2785 & 4.0  & 2696   & 2693 \\
			Li Chao        & 2741 & 4    & 2660 & 2623 & Salem           & 2656 & 3.5  & 2624   & 2592 \\
			Riazantsev     & 2651 & 3.5  & 2641 & 2622 & Tomashevsky     & 2711 & 3.5  & 2629   & 2591 \\
			Hammer         & 2629 & 3    & 2590 & 2562 & Hammer          & 2628 & 3.5  & 2648   & 2617 \\
			Gelfand        & 2719 & 3    & 2582 & 2555 & Riazantsev      & 2671 & 3.0  & 2590   & 2560 \\
			\bottomrule
		\end{tabular}
		\caption{Perfect Performance Ratings in the 2017 FIDE Grand Prix Series in Palma de Mallorca and Sharjah 9-round Swiss tournaments}
		\label{tab:ppr}
		\floatfoot{Note: Rtg = Rating; Pts = Points; TPR = Tournament Performance Rating; PPR = Perfect Performance Rating.}
	\end{table}

	Table~\ref{tab:ppr} presents the PPRs from two top Swiss tournaments: FIDE Grand Prix Series in Palma de Mallorca and Sharjah. If players had entered the tournament with their calculated PPRs as initial ratings, their final scores would be equal to the scores predicted by these ratings. In other words, each player's initial PPR perfectly predicts the player's final score.
	
	In the Palma de Mallorca Grand Prix, although Jakovenko ranked first by the TPR tiebreak, Aronian's PPR is higher due to the strong performances of his opponents in this event. For similar reasons, Nakamura's and Svidler's PPRs are also significantly higher than their TPRs. In the Sharjah Grand Prix, the differences between PPR and TPR can also be observed for Maxime Vachier-Lagrave and Grischuk.
	
	\begin{table}[]
		\centering
		\small
		\begin{tabular}{llll|llll}
			\toprule
			Rank & Name         & Points & PPR  & Rank & Name          & Points & PPR  \\
			\midrule
			1    & Fischer      & 18.5   & 2805 & 13   & Hort          & 11.5   & 2556 \\
			2    & Larsen       & 15     & 2669 & 14   & Ivkov         & 10.5   & 2525 \\
			3    & Geller       & 15     & 2669 & 15   & Suttles       & 10     & 2509 \\
			4    & Huebner       & 15     & 2669 & 16   & Minic         & 10     & 2509 \\
			5    & Taimanov     & 14     & 2636 & 17   & Reshevsky     & 9.5    & 2493 \\
			6    & Uhlmann      & 14     & 2636 & 18   & Matulovi & 9      & 2477 \\
			7    & Portisch     & 13.5   & 2620 & 19   & Addison       & 9      & 2477 \\
			8    & Smyslov      & 13.5   & 2620 & 20   & Filip         & 8.5    & 2460 \\
			9    & Polugaevsky  & 13     & 2604 & 21   & Naranja       & 8.5    & 2460 \\
			10   & Gligori & 13     & 2604 & 22   & Ujtumen       & 8.5    & 2460 \\
			11   & Panno        & 12.5   & 2588 & 23   & Rubinetti     & 5      & 2350 \\
			12   & Mecking      & 12.5   & 2588 & 24   & Jimenez       & 5.5    & 2372 \\
			\bottomrule
		\end{tabular}
		\caption{Palma de Mallorca Interzonal Tournament (round-robin), 1970}
		\label{tab:interzonal}
	\end{table}

	Table~\ref{tab:interzonal} shows the PPRs from the 1970 Palma de Mallorca Interzonal Tournament. The Interzonal events were part of the World Chess Championship cycle and are similar to the current Chess World Cup. Since the first official FIDE Elo list was not published until 1971, players in this tournament did not have official ratings. However, because PPRs do not require pre-tournament ratings, estimating the tournament's average rating is sufficient for calculating PPRs. We calculated the average rating of the participating players based on their published 1971 ratings and used this estimate. The resulting PPRs are presented in the table. Remarkably, Fischer achieves a PPR of 2805, more than 135 Elo points above the second-ranked player.
	
	\section{Concluding remarks}
	
	In this note, we introduced a novel method for measuring player performance in tournaments called Perfect Performance Ratings (PPRs). Unlike the Tournament Performance Rating (TPR), PPRs take into account players' actual performance and achieve perfect consistency between predicted and observed outcomes
	
	While chess served as the primary example in this note, PPRs have potential applications across various domains where Elo-based ratings are commonly used. Future research could apply PPRs to other sports, dating applications, online recommender systems, and evaluations of machine learning models, such as large language models. As \citet{elo1978} suggested regarding TPR, PPRs can similarly function as a robust rating system that can be calculated across predefined time periods.

\end{document}